\documentclass[preprint2]{aastex}
\newcommand{\refer}{\reference}
\begin{document} 
\title{The Intrinsic Shapes of Molecular Cloud Fragments over a Range of Length Scales}
\author{C. E. Jones and Shantanu Basu}
\affil{Department of Physics and Astronomy, University of
Western Ontario, London, Ontario, Canada N6A~3K7}
\email{cjones@io.astro.uwo.ca}
\email{basu@astro.uwo.ca}
\shorttitle{Shapes of Molecular Cloud Fragments}
\shortauthors{Jones \& Basu}
\begin{abstract}

We decipher intrinsic three-dimensional shape distributions of molecular 
clouds, cloud cores, Bok globules, and condensations using 
recently compiled catalogues of observed axis ratios for these objects
mapped in carbon monoxide, ammonia, through optical selection, or in 
continuum dust emission. We apply
statistical techniques to compare assumed intrinsic axis ratio 
distributions with observed projected axis ratio distributions.
Intrinsically triaxial shapes
produce projected distributions which agree with observations.  
Molecular clouds mapped in $^{12}$CO are intrinsically triaxial 
but more nearly prolate
than oblate, while the smaller cloud cores, Bok globules, and 
condensations are also intrinsically triaxial but more nearly oblate 
than prolate.  

\end{abstract}

\keywords{ISM: clouds --- ISM: globules --- ISM: structure ---stars: formation}

\section{Introduction}
Numerous catalogues have now 
compiled properties of hundreds (or even thousands) of molecular clouds or
cloud fragments with a large 
range of sizes, allowing
meaningful statistical analysis of properties of molecular clouds, 
cloud cores, and smaller condensations.
With such data sets, the distribution of apparent projected core axis 
ratio $p$ can be used to 
constrain the intrinsic three-dimensional shapes.  
In a previous paper (Jones, Basu, \& Dubinski 2001; hereafter Paper I), we 
investigated the shapes of molecular cloud cores mapped in NH$_3$ (Jijina, 
Myers, \& Adams 1999)
and cores mapped through optical selection (Lee \& Myers 1999) via 
both statistical and analytical methods. We found that strictly axisymmetric 
prolate or oblate shapes cores 
could not reproduce the observed projected axis ratios and that 
molecular cloud cores were triaxial. 

In this paper, we extend our previous work and conduct a statistical analysis 
of seven
recent data sets which include a wide range of sizes of objects, from  
molecular clouds with effective radius as large as 45 pc (Heyer, Carpenter, 
\& Snell 2001)  
to submillimeter dust continuum maps of condensations with major
axes as small as 2800 AU, or $\sim 0.01$ pc (Motte et al. 2001).
The information obtained about the intrinsic shapes of these objects can yield 
insight into the physical processes which govern their evolution and 
subsequent star formation.  We examine the
$^{12}$CO catalogue of molecular regions in the outer Galaxy 
compiled by Heyer et al. (2001) in \S\ 3.1, cores mapped in
NH$_3$ and C$^{18}$O in \S\ 3.2 (Onishi et al. 1996; Tachihara, Mizuno, \&
Fukui 2000), catalogues of Bok globules (Clemens \& 
Barvainis 1988; Bourke et al. 1995) in  \S\ 3.3, and millimeter and 
submillimeter continuum maps of smaller protostellar condensations (Motte, 
Andr\'{e}, \& Neri 1998; Motte et al. 2001)
in  \S\ 3.4.   A discussion and summary are given in \S\S\ 4 and 5.

\section{Methods}
In general, a triaxial ellipsoid can be described by the equation
\begin{equation}
x^2 + \frac{y^2}{\zeta^2} + \frac{z^2}{\xi^2} = a^2,
\end{equation}
where $a$ is a constant and $1 \ge \zeta \ge \xi$.  
The geometrical analysis of Stark (1977) and Binney (1985) shows that such
a body, when viewed in projection, has elliptical contours. Following 
Binney (1985), the projection of a triaxial body when viewed from an
observing angle $(\theta, \phi)$ (where the
angles are defined on an imaginary viewing sphere and have their usual
meaning in a spherical coordinate system) is found using the quantities
\begin{equation}
A \equiv \frac{{\cos}^2 \theta}{\xi^2}\left({\sin}^2\phi +
\frac{{\cos}^2\phi}{\zeta^2}\right) + \frac{{\sin}^2\theta}{\zeta^2},
\end{equation}
\begin{equation}
B \equiv \cos \theta \, \sin2\phi \left(1 - \frac{1}{\zeta^2}\right) \frac{1}{\xi^2},
\end{equation}
and
\begin{equation}
C \equiv \left(\frac{{\sin}^2 \phi}{\zeta^2} + {\cos}^2\phi \right)
\frac{1}{\xi^2}.
\end{equation}
The apparent axis ratio in projection then equals
\begin{equation}
p = {\left(\frac{A + C - D}{A + C + D}\right)}^{1/2},
\label{binneyform}
\end{equation}
where $D \equiv \sqrt{{\left(A - C \right)}^2 + B^2}$. 
Using these equations, one can construct probability distributions
for the projected axis ratio, assuming a large
number of randomly distributed viewing angles.

We assign a Gaussian distribution of values
for each axis ratio $\zeta$ and $\xi$, with a mean in the range $[0,1]$, 
and standard deviation $\sigma$ typically equal to 0.1, consistent
with our use of 10 bins to sample the data. 
We did test a range of $\sigma$ from 
0.05 to 0.2, and find that our conclusions do not change significantly within 
this range (see discussion in \S\ 4). 
The drawback to using $\sigma \gtrsim 0.2$ 
is that a relatively large fraction of the Gaussian distribution can fall
outside the allowed range $[0,1]$. For example, a Gaussian
distribution centered at 0.8 with a $\sigma$ of 0.2 has $\approx$ 16\%
of the data $>1$ (see Paper I for an extended discussion 
of this issue). 
For similar reasons, we limit the $\sigma
= 0.1$ analysis to the range of axis ratios [0.1, 0.9].
%We tested several different ways to deal with these values:
%(1) we set values less than zero to zero and values greater
%than one to one; (2) we removed all numbers outside of the
%range of zero to one; (3) we rejected numbers which
%fell outside of the range of zero to one and repeatedly generated a
%new random number until one between zero and one was obtained.
%None of these methods is completely
%satisfactory since Gaussian distributions modified by these methods
%clearly have different means and standard deviations than originally
%specified.  We believe that it is not very meaningful to test Gaussian
%distributions with larger values of $\sigma$ than 0.2 especially near
%the end-points of our parameter space.  For example, a Gaussian
%distribution centered at 0.8 with a $\sigma$ of 0.2 has $\approx$ 16\%
%of the data $>1$. If one simply truncates the values
%$>1$, the new resultant distribution is centered at 0.7 with a
%standard deviation of 0.16.  

In order to find the best fit intrinsic distribution of triaxial bodies, 
distributions of axis ratios with peak values $\xi_0$ and $\zeta_0$ 
(for a given $\sigma$) are
input into a Monte Carlo program. We typically 
employ at least $10^4$ viewing angles to calculate the projected distribution
for each individual pair of axis ratios, and at least $10^4$ sets of axis 
ratios for each Gaussian distribution. This is more than sufficient for 
comparison with data sets sampled in 10 bins. The program
produces the expected observed distribution which results from the assumed
intrinsic distributions. We compare this output with the observed data sets
via their $\chi^2$ values calculated by comparing the area of each bin to the 
area under 
the expected distribution at the location of each bin.  Distributions of
triaxial shapes for 
which the mean values $\xi_0 = \zeta_0$ emphasize prolate objects, while those
distributions with $\zeta_0$ near 1 emphasize oblate objects. Furthermore,
distributions with $\zeta_0 > \onehalf (1 + \xi_0)$ can be classified as 
containing more nearly oblate than prolate objects.
For greater details consult Paper I.

\section{Results}
\subsection{Molecular Clouds}

Heyer et al. (2001) have catalogued the properties of clouds and clumps using
$^{12}$CO data from a survey of the
outer Galaxy.  The catalogue consists of $10\,156$ 
objects which include small, isolated clouds and clumps within larger clouds.
Observations from the outer Galaxy have the advantage that there is no 
distance ambiguity and the clouds are 
more widely spaced along the line of sight which eliminates problematic 
blending of emission, allowing cloud properties to be determined more 
accurately.  The FWHM 
beam size for this catalogue is 
$45^{'\!'}$ and there are an impressive $10\,134$ objects with major and minor 
axes tabulated. Heyer et al. (2001) point out that the
largest clouds, with effective radius $r_{\rm e}$ of about 10 pc or larger, are 
self-gravitating, whereas the smaller clouds (comprising the vast majority 
of observed objects) are not self-gravitating. This boundary also 
approximates the usual distinction between molecular clouds and giant 
molecular clouds, or GMC's (see Blitz 1991; Williams, Blitz, \& McKee 2001).

Given the evidence for a physical distinction based on size,
we performed our analysis on the entire data set as well as a 
subset with $r_{\rm e} > 10$ pc, which corresponds to a mass range of
approximately
$\sim 10^4\: \rm{to}\: 10^5 \:\rm{M}_{\odot}$, typical of GMC's.
Furthermore, to avoid any pathological cases where an elliptical fit to the 
cloud shape can be a very poor approximation (M. H. Heyer, 2001, 
private communication), we restricted our sample to those clouds
which span at least 10 spatial pixels in the observations (an even more
restrictive threshold of 20 pixels also yields essentially the same 
final result).
This criterion reduces the total set to 5685 and the subset of GMC's to 85.
Our separate analyses can reveal whether there is any significant 
shape difference in the two populations.

\begin{figure}
\epsscale{1.0}
\plotone{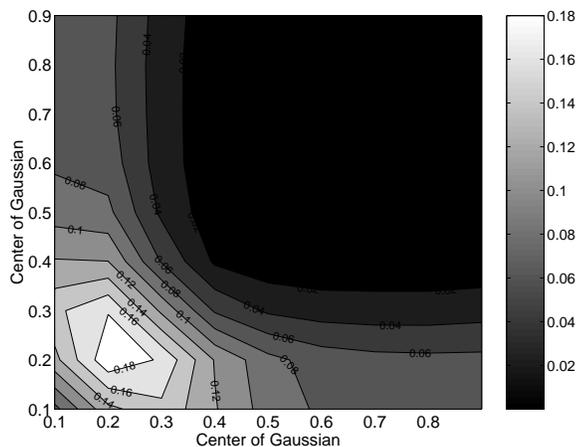}
\caption{Two-dimensional plot of inverse $\chi^2$ values for triaxial core 
shape models applied to clouds with effective radius greater than 10 pc (and 
observed with at least 10 pixels) in the catalogue of Heyer et al. (2001).  Note the symmetry of the figure about the line along which the two centers of the axis distributions are equal.  For any point in the figure, the smaller axis ratio corresponds to $\xi_0$ and the larger to
$\zeta_0$. }
\label{heyerlarge2d}
\end{figure}

Figure~\ref{heyerlarge2d} shows
the results of the $\chi^2$ calculations for the triaxial fitting of the 
GMC subset.  The data 
set is best fit by distributions with axis ratios 
$(\xi_0,\zeta_0) = (0.2,0.2)$ when $\sigma=0.1$.  In order to determine 
whether values of  $(\xi_0,\zeta_0)$ closer to zero would improve the fit, we repeated the 
analysis with a value 
of $\sigma=0.05$.
(As the 
mean value of the Gaussian gets 
closer to the endpoints of the allowed range [0,1], a larger fraction of the 
Gaussian distribution falls outside this range, for a given $\sigma$.  There is no ideal way to 
correct for this problem as explained in \S\ 2 of this paper and in \S\ 4 of 
Paper I).  However, even with
$\sigma=0.05$, the best fit mean axis ratios $(\xi_0,\zeta_0)$ are not closer to zero and agree with our result for  $\sigma=0.1$ within the estimated error.  
See \S\ 4.1 for a
discussion of the errors which we estimate to be a maximum of 
$\pm \, 0.1$ in the mean value of each axis ratio.  

\begin{figure}
\epsscale{1.0}
\plotone{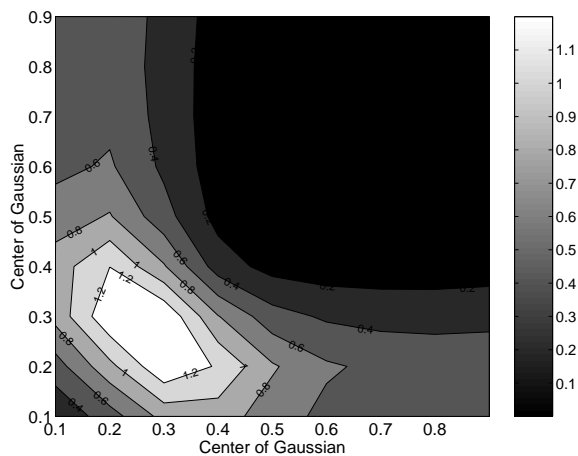}
\caption{Two-dimensional plot of inverse $\chi^2$ values for triaxial core 
shape models applied to the entire catalogue of Heyer et al. (2001) based
on our selection criterion.}
\label{heyertotal2d}
\end{figure}

Figure~\ref{heyertotal2d} shows the result of the $\chi^2$ calculation for the
complete set of axis ratios based on our selection criteria in the 
Heyer et al. (2001) catalogue.  
The best fit based on the $\chi^2$ values is 
$(\xi_0,\zeta_0) = (0.3,0.3)$.  

For both the total set and for the subset based on large effective radius, 
the best-fit triaxial
distributions require $\xi_0 = \zeta_0$, as
shown very clearly in Figure~\ref{heyerlarge2d} and 
Figure~\ref{heyertotal2d}.  This means that the distributions 
emphasize prolate objects. However, since $\xi_0$ and $\zeta_0$ are the means 
of {\it distributions}, most individual objects cannot be considered strictly 
prolate and are, in
fact, triaxial.  For the clouds with large $r_{\rm e}$, the 
distributions which best
fit the observations have thinner objects (smaller $\xi$ and $\zeta$)
than those that best fit the entire set.
However, the difference is quite small and equal to our maximum estimated 
error of $\pm\, 0.1$.

\begin{figure}
\epsscale{1.0}
\plotone{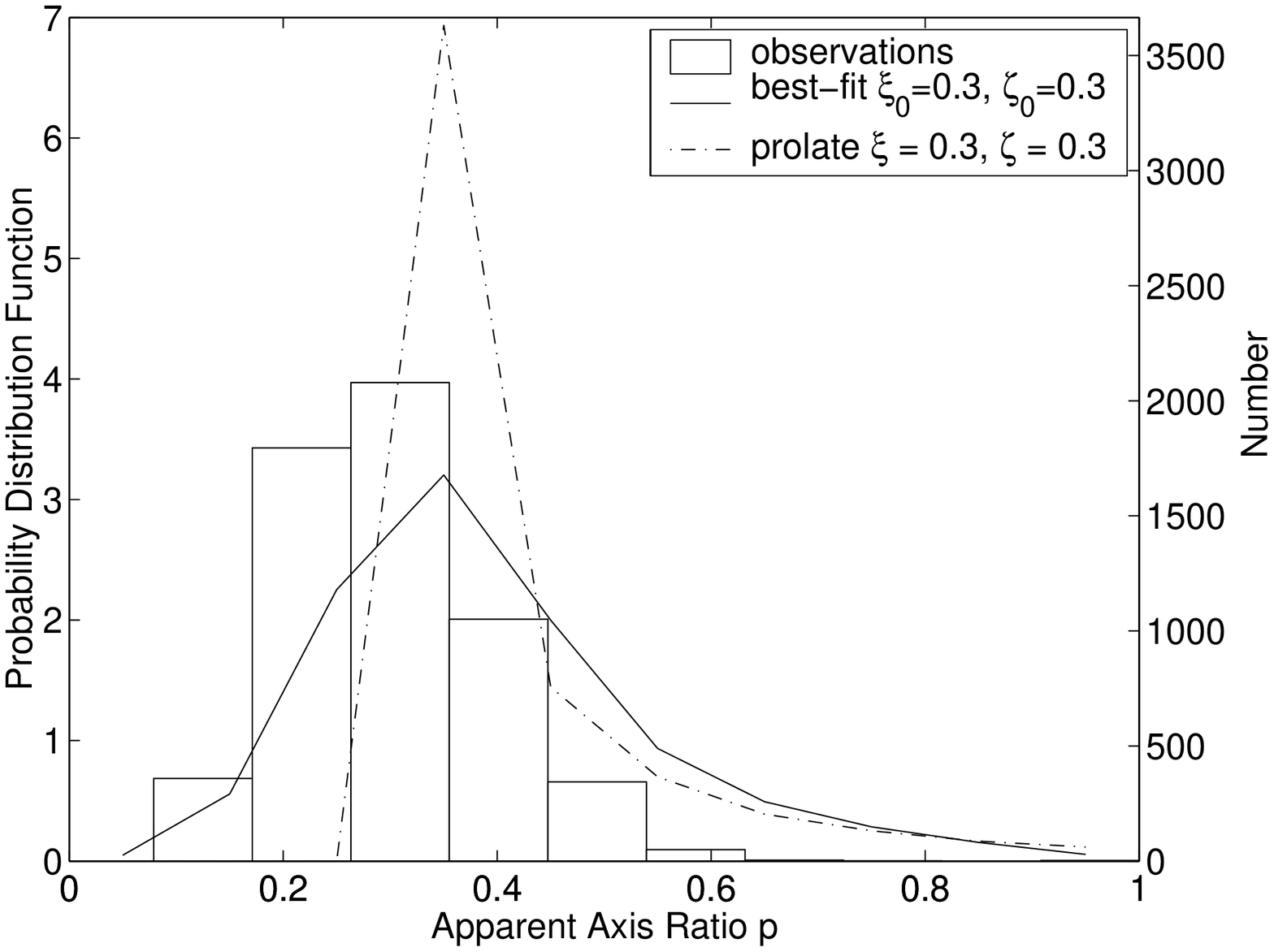}
\caption{Comparison of the observed axis ratios from Heyer et al. (2001) with 
the best fit assuming triaxial clouds (solid line) and assuming a pure prolate 
cloud (dot-dashed line).  The actual number of observations in the bins is 
shown on the right-hand side vertical axis.}
\label{heyertotalbest}
\end{figure}

Figure~\ref{heyertotalbest} compares the best fit distribution of $p$
to the complete binned
data set of Heyer et al. (2001). It also reveals that the histogram of 
projected shapes 
$p$ of molecular clouds has some unique features.  We recall that our 
previous analysis (Paper I) of dense cores showed that an observed
broad peak in the distribution at $p \gtrsim 0.5$ and the presence of a 
significant number of objects near 
$p=1$ favored triaxial, but more nearly oblate intrinsic shapes. Indeed,
this pattern is reinforced in our subsequent study of other cores, Bok globules,
and protostellar condensations (see \S\ 3.2 - 3.4).
However, the shapes of molecular clouds are distinct in that they have
a very narrow peak, and at a low value $p \approx 0.3$. The narrow peak favors
near-prolate objects, although a pure prolate cloud with $\xi=\zeta=0.3$
yields a poor fit to the data, as shown in Figure~\ref{heyertotalbest}.
A pure prolate cloud would have too narrow a peak in the observed shape
distribution, as well as a higher probability than observed of a 
near-circular projection (see discussion and figures in \S\ 2 of Paper I). In 
fact,
Figure~\ref{heyertotalbest} shows that even
a Gaussian distribution of triaxial objects implies a higher probability
of detecting near-circular objects than observed. 

The cutoff in high values of $p$ may be due to 
an actual cutoff in the distribution of intrinsic axis ratios 
$(\xi_0,\zeta_0)$ above some value, or could be due to some selection effect. 
One selection effect in the 
Heyer et al. (2001) sample is the fact that an object must span 
at least 5 pixels in the map to be classified as a cloud (additionally,
we impose the higher threshold of 10 pixels for our shape analysis).
However, we see no evidence in the sample for a trend toward greater 
circularity as clouds have smaller projected size. Another selection 
effect is that the edges of the objects in the sample likely correspond to 
the the CO photodissociation boundary and not that of the H$_2$ gas
(Heyer et al. 2001). However, it is again not clear that this in any way
biases against near-circular objects.

%The method that Heyer et al. (2001) uses to determine the major and 
%minor axes is quite simple.  The major 
%axis is the chord which connects the 2 most extreme points in the cloud and 
%the minor axis forms an equivalent area with the number of pixels. Although,
%this technique may hide complex features within the cloud, we do not know of 
%any selection effects which reduce the near-circular projections (Heyer 
%2001).  Perhaps there 
%is an abrupt cutoff in intrinsic axis ratios $(\xi_0,\zeta_0)$ above some 
%value.

\subsection{Dense Cores}

\begin{figure}
\epsscale{1.0}
\plotone{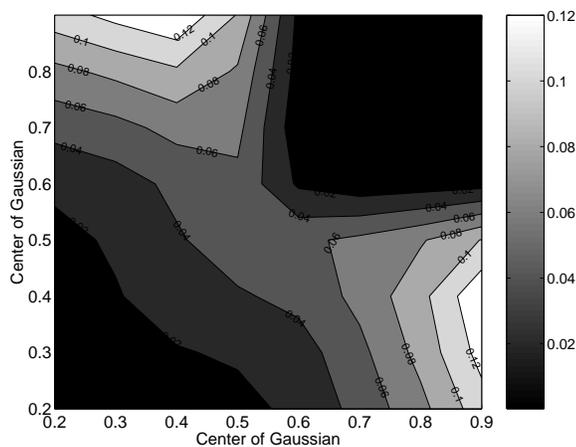}
\caption{Two-dimensional plot of inverse $\chi^2$ values for triaxial core 
shape models applied to the combined surveys of Onishi et al. (1996) and 
Tachihara et al. (2000).}
\label{japtri}
\end{figure}

\begin{figure}
\epsscale{1.0}
\plotone{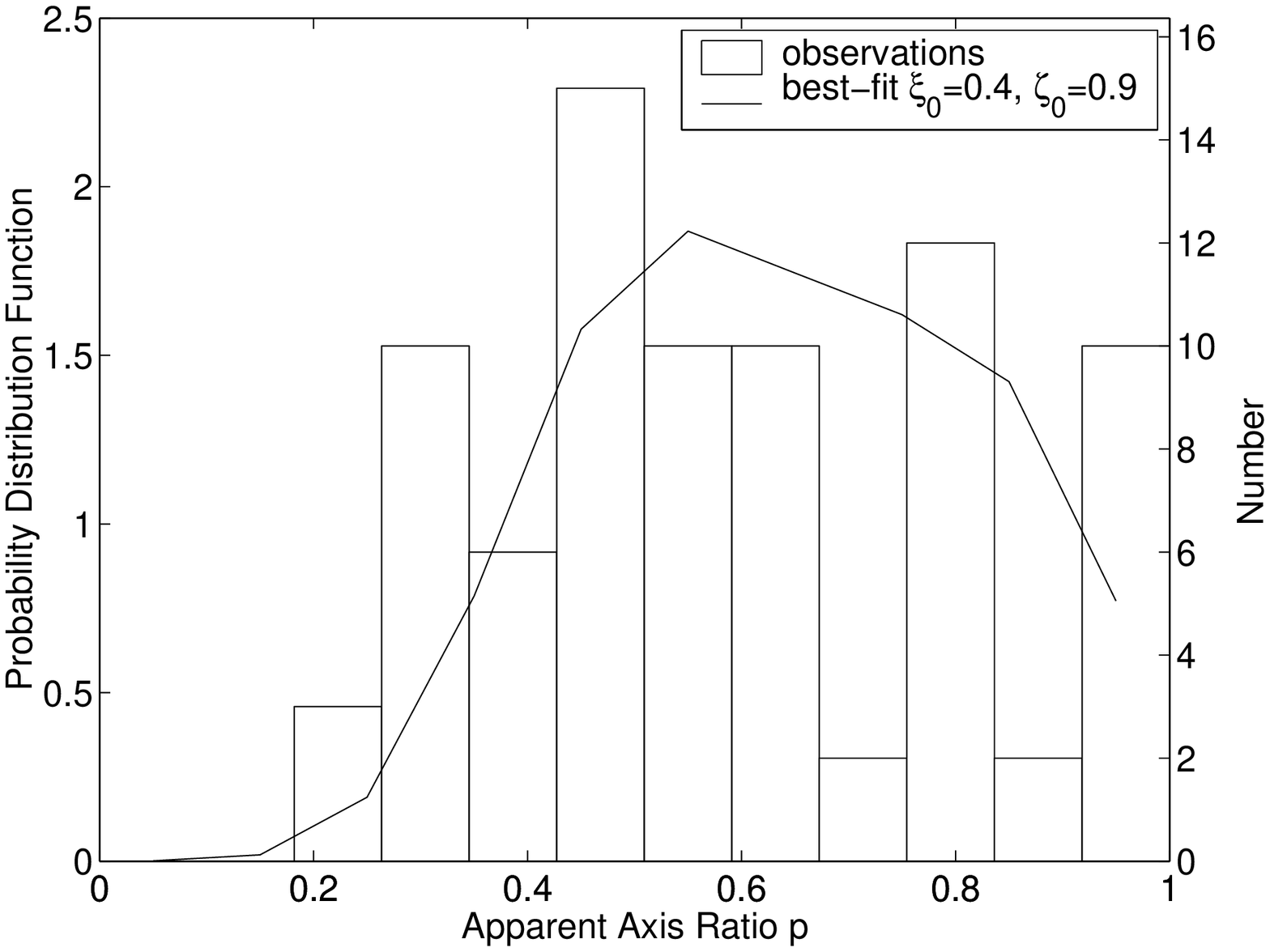}
\caption{Comparison of the observed axis ratios from Onishi et al. (1996) and 
Tachihara et al. (2000) with 
the best fit assuming triaxial cores (solid line).  The actual number of observations in the bins is shown on the  right-hand side vertical axis.}
\label{bestjap}
\end{figure}

Onishi et al. (1996) and Tachihara et al. (2000) have observed dense cloud 
cores in Taurus and Ophiuchus, respectively, in C$^{18}$O with 
telescopes at Nagoya University.  
Since the same telescopes and 
technique were used to obtain these ratios, and both groups quote the axial 
ratios to a tenth of a parsec, we combined these two surveys in 
order to obtain a reasonable sized set for statistical analysis.  
The final set of data from these two regions 
includes 80 cores.  The best fit determined by 
calculating the $\chi^2$ 
values is $(\xi_0,\zeta_0) = (0.4,0.9)$; see Figure~\ref{japtri}. 
The best fit is compared with the 
binned data set in Figure~\ref{bestjap}.  
We performed this analysis with and without the sources for which there were
embedded IR sources and obtained the same result.  

Although our best fit 
underestimates the observed axis ratios in the bin 
near $p = 1$, it also overestimates the previous bin.  The significant bin to 
bin variation is likely due to the relatively small size of this sample.  
However, the best fit agrees within the estimated errors, with the 
results presented in Paper I using much larger samples of NH$_3$ data 
(Jijina et al. 1999) and optical data (Lee \& Myers 1999) for dense cores.  
Jijina et al. (1999) catalogued core properties for 264 objects from NH$_3$ 
observations and Lee \& Myers (1999) catalogued properties for 406 dense 
cores from contour maps of optical extinction.  The Jijina et 
al. data set is best fit by 
triaxial distributions with mean
axis ratios $(\xi_0,\zeta_0) = (0.5,0.9)$, and the Lee \& Myers data 
set is best fit by mean values $(\xi_0,\zeta_0) = (0.3,0.9)$.

\subsection{Bok Globules}

\begin{figure}
\epsscale{1.0}
\plotone{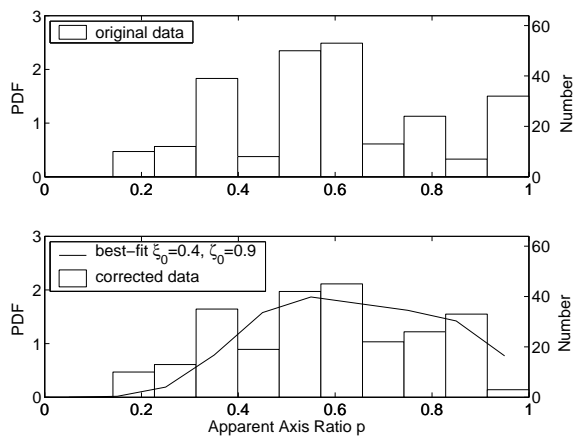}
\caption{Clemens \& Barvainis (1988) original binned normalized data set and 
the same data 
corrected for rounding errors.  The original and 
corrected data are displayed in the top and bottom frames, respectively.  
The black solid line is the best fit from the
triaxial distributions.  The actual number of objects in the bins is 
shown on the right-hand side vertical axis.}
\label{clemoriandf.eps} 
\end{figure}

\begin{figure}
\epsscale{1.0}
\plotone{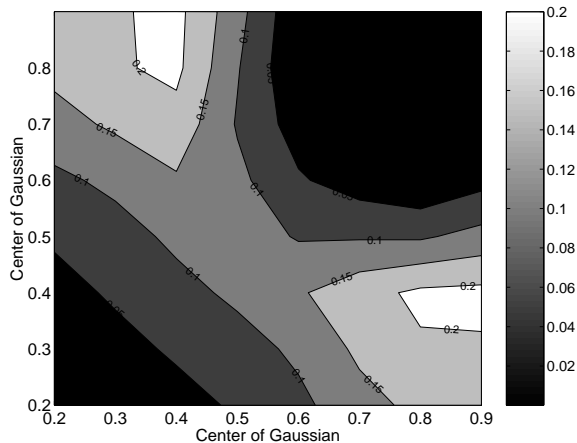}
\caption{Two-dimensional plot of inverse $\chi^2$ values for triaxial core 
shape models applied to the catalogue of Clemens \& Barvainis (1988) with the 
rounding errors corrected.} 
\label{clemtrifig.eps}
\end{figure}

\begin{figure}
\epsscale{1.0}
\plotone{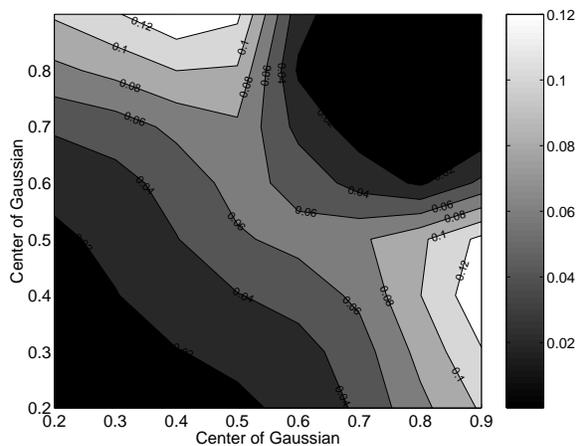}
\caption{Two-dimensional plot of inverse $\chi^2$ values for triaxial core shape models applied to the catalogue of Bourke et al. (1995) with the 
rounding errors corrected.}
\label{bourketrifit.eps}
\end{figure}

Bourke et al. (1995) catalogued physical characteristics of 169 isolated small
molecular clouds (Bok globules) in NH$_3$ from the southern sky.  These 
observations were compiled to complement a similar study of 248 optically 
selected Bok globules in the northern hemisphere by Clemens \& 
Barvainis (1988).  We note that Clemens \& Barvainis (1988) also found that
there was no correlation between the orientation of the clouds relative to
the Galactic plane
and their projected optical shapes.
Ryden (1996) looked at the shapes of the globules from 
these two data sets based on the assumption of axisymmetry.  She found that
the Bok globules were consistent with oblate objects having an intrinsic 
mean axis ratio $q=0.3$, or prolate objects having $q=0.5$.
Ryden (1996) realized that these sets suffered from
rounding errors.  This affected the smallest major and minor axes,
yielding an artificially large number of axis ratios near one.  For example,
Clemens \& Barvainis (1988) measured the Palomar Observatory Sky Survey 
plates to compile their catalogue.  They rounded the major and minor axes 
to the nearest millimeter, which corresponds to an angular size of 
$1^{'}\!.12$.  The set is comprised of a relatively large number of small 
globules (42 out of 248) which have a major axis less than 2 mm.  When these
smallest values are rounded, the
result is a large number of axis ratios erroneously equal to one or close to 
one.  The top panel of 
Figure~\ref{clemoriandf.eps} shows very clearly that there are proportionally 
a large number number of objects in the bin with the largest axis ratios.
Obviously, this problem does not affect the larger 
globules as significantly.  Ryden (1996)
compensated for this problem by adding to {\it each} of the 
original major and minor axes a random error term $\Delta$.  She did this 
hundreds of times and obtained a new distribution of major and minor 
axes by taking an average from all her trials.  See Ryden (1996) for 
additional details.  We followed her procedure to 
correct for these rounding errors and obtained a new 
distribution of axis ratios for both data sets.  
Since the roundoff error is likely selected uniformly from within a fixed 
range, we used a random number generator to select uniformly from a 
prescribed range of estimated error. For the
Clemens \& Barvainis (1988) and the Bourke et al. (1995) data sets,
we used a range of error
$-0^{'}\!.56 < \Delta < +0^{'}\!.56$, and 
$-0^{'}\!.25 < \Delta < +0^{'}\!.25$, respectively. 
Figure~\ref{clemoriandf.eps} shows both the original binned data of axis 
ratios and the corrected distribution of axis ratios for the 
Clemens \& Barvainis (1988) data.  Notice that the large
number of axis ratios near one in the original set has been dramatically
reduced.  We investigated these two sets and found that intrinsic triaxial
shapes produced distributions which
agreed with the observations.  Figure~\ref{clemtrifig.eps} and 
Figure~\ref{bourketrifit.eps} show
the results of the $\chi^2$ calculations with the triaxial fitting for the 
Clemens \& Barvainis (1988) and the Bourke et al. (1995) corrected data sets, 
respectively. 
The Clemens \& Barvainis (1988) data set is best fit by 
distributions with mean
axis ratios $(\xi_0,\zeta_0) = (0.4,0.9)$, and the Bourke et al. (1995) data 
set is also best fit by mean values $(\xi_0,\zeta_0) = (0.4,0.9)$.  The best 
fit for the Clemens \& Barvainis (1988) data is also shown in 
Figure~\ref{clemoriandf.eps}.

\subsection{Condensations Mapped in Millimeter-Submillimeter Wavelengths}

\begin{figure}
\epsscale{1.0}
\plotone{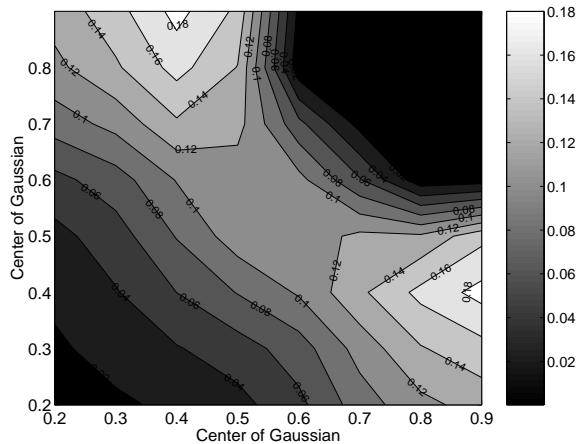}
\caption{Two-dimensional plot of inverse $\chi^2$ values for triaxial core 
shape models applied to the survey of Motte et al. (1998) with the composite clumps
removed.}
\label{trimandn}
\end{figure}

Motte et al. (1998) observed the $\rho$ Ophiuchi main cloud at 1.3 mm with 
the IRAM 30 m telescope and were able to detect 58 individual compact dusty 
objects with a fragmentation size scale of approximately 6000 AU.  After we 
removed clumps which were thought to be composite in nature, there were 
35 clumps for which projected major and minor axes were resolved.  Although 
this is a small sample for statistical analysis, we still present the    
results of the $\chi^2$ calculations from the triaxial fitting in 
Figure~\ref{trimandn}.  The Motte et al. (1998) data set is best fit by 
distributions with mean axis ratios $(\xi_0,\zeta_0) = (0.4,0.9)$.

\begin{figure}
\epsscale{1.0}
\plotone{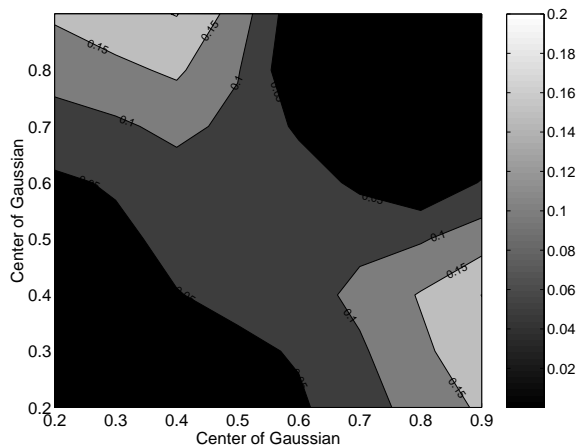}
\caption{Two-dimensional plot of inverse $\chi^2$ values for triaxial core 
shape models with $\sigma = 0.1$ applied to the survey of Motte et al. (2001).}
\label{trimotte1sig}
\end{figure}

\begin{figure}
\epsscale{1.0}
\plotone{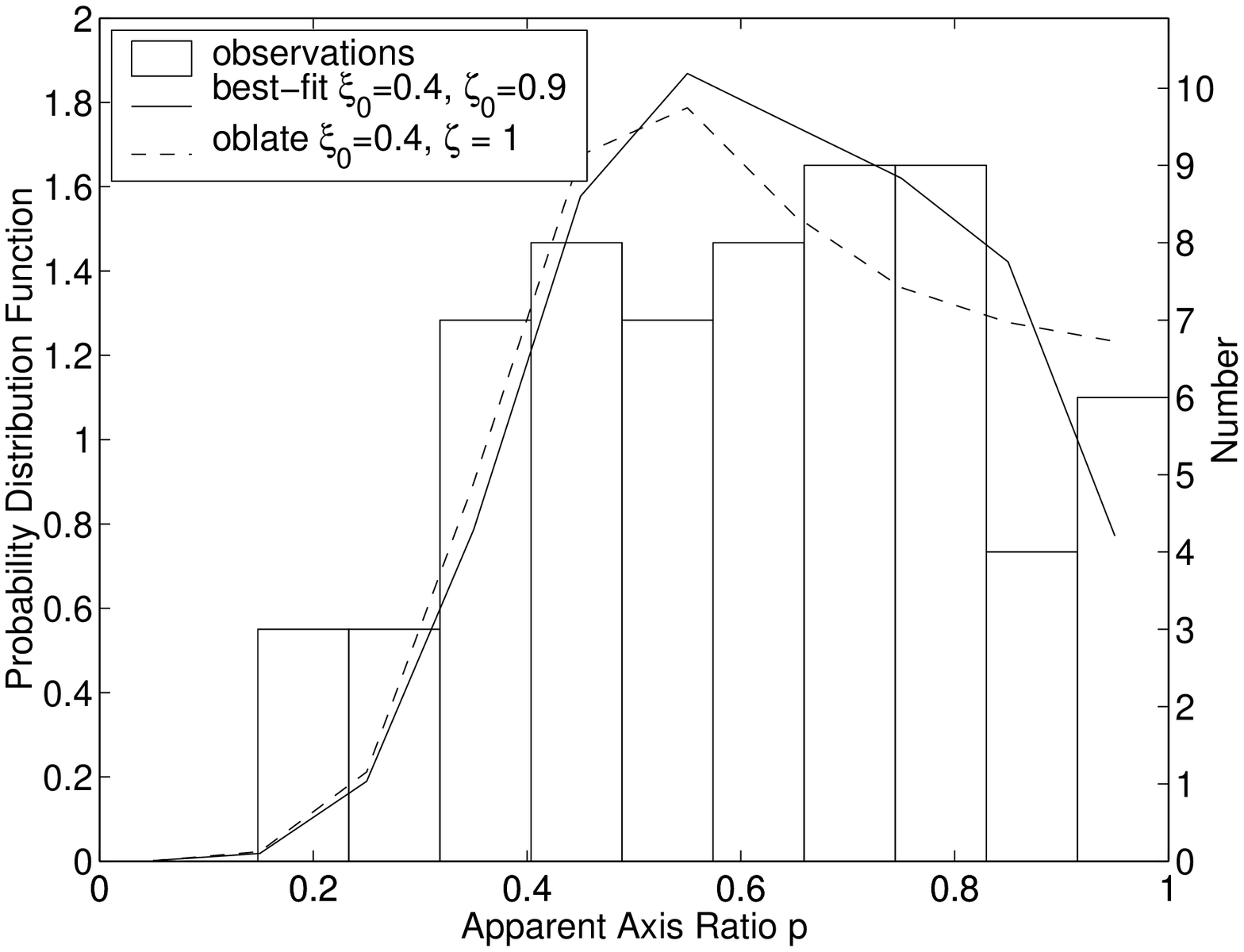}
\caption{Comparison of the observed axis ratios from Motte et al. (2001) with 
the best fit assuming triaxial cores for $\sigma = 0.1$ (solid line) and
assuming pure oblate cores (dashed line).  The 
actual number of observations in the bins is shown 
by the right-hand side vertical axis.}
\label{mottebest}
\end{figure}

Motte et al. (2001) surveyed the protoclusters NGC 2068 and NGC 2071 in Orion
B at 850 $\mu$m and 450 $\mu$m with SCUBA (Submillimeter Common User 
Bolometer Array) on the JCMT and were able to detect
small objects (size $\sim$ 5000 AU) which they call condensations.
There were 64 condensations for which the projected major and minor axes
were obtained. This is a remarkable data set which measures objects on
size scales approaching that of our solar system.  
The results of the $\chi^2$ calculations with the triaxial fitting are 
presented in 
Figure~\ref{trimotte1sig} for $\sigma = 0.1$.  The Motte et al. (2001) data 
set is best fit by 
distributions with mean axis ratios $(\xi_0,\zeta_0) = (0.4,0.9)$.  Since the 
binned observed data of axis ratios does not have a strong peak, but instead 
looks rather flat, we 
repeated our analysis for $\sigma = 0.2$.  
For $\sigma = 0.2$, the data set is best fit by 
distributions with mean axis ratios $(\xi_0,\zeta_0) = (0.5,0.9)$.  However, 
the 
distribution with the mean value of 0.9 and $\sigma = 0.2$, has $31\%$ of the 
data values outside the allowed range.  After the data 
is truncated, the remaining distribution will not have a final mean at 0.9 and 
$\sigma = 0.2$, and 
the remaining values inside the allowed range must also be 
normalized.  As a result, the normalized distribution is actually peaked 
more than one would intuitively expect.  Consequently, the $\chi^{2}$ 
values for $\sigma = 0.2$ are no better than the $\chi^{2}$ values for 
$\sigma = 0.1$.  The best 
fit for $\sigma = 0.1$ is compared with the observed data in
Figure~\ref{mottebest}.  

Interestingly, Figure~\ref{mottebest} shows
that there are a number of objects with an apparent axis
ratio near or exactly equal to one.  A close examination of the 
original Motte et al.
(2001) data reveals that the objects which fall into this bin are well above
the resolution limit and that this large number is not a selection effect.
Since this data set has a number of axis ratios near $p=1$,
it is interesting to speculate whether or
not intrinsic oblate objects (rather than near-oblate triaxial objects)
may agree with the observations. 
We used two additional tests to investigate this hypothesis.
First, we fit the histogram of observed shapes with a curve and carry out
an analytic inversion to see if it is consistent with a distribution 
of intrinsic
oblate shapes. The resulting intrinsic shape distribution $\psi(q)$
does go negative near $q=1$ just as in the inversions performed on the
dense core shape data of Jijina et al. (1999) and Lee \& Myers (1999) 
in Paper I. This is due to the continuing
presence of a decline in the observed axis ratio distribution 
$\phi(p)$ towards $p=1$.
To investigate further, we also experiment with an intrinsic axis 
distribution that is Gaussian in $\xi$, with $\xi_0 = 0.4$ and $\sigma = 0.1$,
but has a
fixed value $\zeta = 1$ for all clouds, i.e., the clouds are all oblate
with various degrees of flattening.
In this case, our Monte Carlo program yields a probability distribution 
function which is plotted in Figure~\ref{mottebest}. The inverse 
$\chi^2$ values for both the pure oblate objects and the best fit triaxial 
objects are $\approx 0.2$. The Motte et al. (2001) data, unlike the dense
core data presented or reviewed in \S\ 3.2, is close enough to being 
flat near $p=1$ and has a small enough
sample of objects that we cannot entirely exclude the pure oblate hypothesis 
on the basis of the $\chi^2$ test.
If just one of the six objects (interestingly, one object has an embedded 
class 0 
protostar) is removed from the bin closest to $p=1$, the near-oblate triaxial
objects provide a much better fit. 
Further observations and a larger sample will be necessary before the 
intrinsic shapes of these very
dense objects can be determined with certainty.  
Nevertheless, we can conclude that molecular cloud 
condensations are either oblate or near-oblate in shape.

%\begin{figure}
%\epsscale{1.0}
%\plotone{f11.eps}
%\caption{The same as figure~\ref{trimotte1sig} except with $\sigma = 0.2$.}
%\label{trimotte2sig}
%\end{figure} 

%In order to test the robustness of our result, we remove 20\% of the data in
%each sample randomly and recalculate the
%reduced $\chi^2$ values.  This is repeated
%10 times for each data set.  Each such simulation with the Jijina et al.
%(1999) data set continues to have a best fit at mean axis ratios 0.5 and 0.9.
%The Lee \& Myers
%data has a best fit with one mean axis ratio 0.3 in all cases, and the
%other in the range 0.7 to 0.9, with a mean of 0.83.

\section{Discussion}

\subsection{Error Bounds}

In order to test
the reliability of our results we recalculate the $\chi^2$ values after
randomly removing $20\%$ of the data and repeat this procedure ten times for
each set.  For every set investigated in this paper we obtain the same best
fit mean axis ratios $(\xi_0,\zeta_0)$ for all the trials with the
exception of two trials with the Motte et al. (1998) data. This data set
has only 35 non-composite objects for which the minor and major axis are
published.  When $20\%$ of the data is randomly removed, only 28 objects remain.
Nevertheless, we obtain $(\xi_0,\zeta_0) = (0.4,0.9)$ for eight trials,
$(\xi_0,\zeta_0) = (0.5,0.9)$ for one trial, and $(\xi_0,\zeta_0) = (0.4,0.8)$
for the remaining trial.

Additionally, adjusting the width of the Gaussian distributions in 
$\xi$ and $\zeta$ within the range $[0.05,0.2]$ yields at most a change
of $\pm 0.1$ in the best fit values $\xi_0$ and $\zeta_0$. 
The largest variation is seen in the smallest data sets. The effect on 
the Motte et al. (2001) data is described in \S\ 3.4. Using $\sigma =0.05$
allows us to test distributions with peaks near the boundaries 0 and 1, 
while a relatively large $\sigma=0.2$ allows us to see if a better fit
exists for the broadest observed distributions in projected axis ratio $p$.

Altogether, our testing allows us to state an approximate maximum error in
our best fit mean axis ratios of $\pm 0.1$.  Interestingly, absolutely 
{\em all} of the best fits for the data sets of objects with size 
scale $\sim 0.1$ pc or smaller agree with one another 
within this range of error. 
This strengthens the case that these objects are triaxial but 
preferentially flattened in one direction and close to oblateness.  
A small degree of triaxiality seems necessary to explain the observed
decline near $p=1$, and the pure oblate hypothesis seems excluded
for dense cores (Paper I and \S\ 3.2).
Due to the smaller number statistics and relatively large number of 
objects with $p$ near one, the dense condensations are most likely
to be compatible with pure oblateness. Future more extensive observations
of this class of objects is necessary to settle the issue.

Finally, there is the possibility that our results are biased
by the fact
that the spectral line or dust continuum data are probing
emission from
regions of varying opacity and/or temperature, so that the
projected
shape may not correspond exactly to the physical shape of the
gas
distribution. When data from other wavelengths become available,
it is
possible our conclusions may change. This applies particularly
to the
molecular cloud data (\S\ 3.1), for which we have used a single
catalogue.
However, since we obtained very nearly the same result using a
variety of
tracers for the smaller cloud cores, Bok globules, and
condensations,   
the results in these cases seem robust.

\subsection{Physical Implications}

This investigation shows that intrinsic triaxial objects produce 
distributions which reasonably match observations of projected axis ratios for
molecular clouds, molecular cloud cores, Bok globules, and 
protostellar condensations. The results clearly fall into two categories:
(1) on scales $\gtrsim 1$ pc, mapped in $^{12}$CO, molecular
clouds, including GMC's, have triaxial shapes which are more closely
prolate than oblate; (2) dense cores, Bok globules, and 
condensations, mapped in a variety of tracers, on scales from 
few $\times \,\, 0.1$ pc down to $0.01$ pc, have triaxial shapes which are 
more closely oblate than prolate.
The results about the latter objects reinforce our earlier finding (Paper I) 
from two other catalogues of dense core shapes.  
See Table~\ref{summary} for a summary of the best fit axis ratios, 
($\xi_0, \zeta_0$), for each
of the data sets we investigated.  The results from 
Paper I are included for comparison.

\begin{deluxetable}{lcrr}
\tablecolumns{4} 
\tablewidth{0pc} 
\tablecaption{Summary of Best Fit Mean Axis Ratios} 
\tablehead{ 
\colhead{Data Set} & \colhead{Object Type}   & \colhead{$\xi_0$}    & \colhead{$\zeta_0$}\\
\colhead{} & \colhead{}& \colhead{$\pm \, 0.1$}& \colhead{$\pm \, 0.1$} }
\startdata 
Heyer et al. (2001)&clouds with effective radius $>$ 10 pc & 0.2& 0.2\\ 
&complete set of clouds& 0.3& 0.3\\
\cline{1-4}\\ 
Onishi et al. (1996) \& &molecular cloud cores& 0.4& 0.9\\
Tachihara et al. (2000)&  & &\\
Jijina et al. (1999)\tablenotemark{a}& molecular cloud cores & 0.5& 0.9\\
Lee \& Myers (1999)\tablenotemark{a}& molecular cloud cores & 0.3& 0.9\\
Clemens \& Barvainis (1998)& Bok globules & 0.4& 0.9\\
Bourke et al. (1995)& Bok globules & 0.4& 0.9\\
Motte et al. (1998) & dense condensations & 0.4& 0.9\\
Motte et al. (2001) & dense condensations & 0.4& 0.9\\
\enddata 
\label{summary}
\tablenotetext{a}{Previous result from Paper I}
\end{deluxetable}

The robust tendency for cores, Bok globules, and smaller condensations to have
triaxial fits with $\xi_0 = 0.3-0.5$, and $\zeta_0 = 0.9$ implies that
they are all preferentially flattened in {\em one} direction.  
This could be due to flattening along the direction of a mean magnetic
field, or due to significant rotational support in the smallest objects.
The magnetic field explanation for cores is attractive as it implies that the
observed near-alignment of core minor axes and magnetic field direction
in Taurus (see  Onishi et al. 1996) may be indicative of a more universal
phenomenon. We also note that early submillimeter polarimetry of a few 
dense cores (Ward-Thompson et al. 2000) reveals a tendency toward 
alignment, 
but also a noticeable angular offset. This is interpreted as evidence 
for triaxiality of the cores (Basu 2000), which may still be preferentially 
flattened along the direction of the magnetic field. 

While triaxiality is consistent with a nonequilibrium state, evolving due
to external turbulence or internal gravity, the near-oblate shape also 
means that the objects may not be particularly far from equilibrium, and
that oblate equilibrium models may act as a reasonable approximation to
these objects. This can explain why the internal structure of some Bok
globules and pre-stellar cores can be closely or approximately fit by
spherical equilibrium Bonnor-Ebert or near-equilibrium oblate magnetic models
(Alves, Lada, \& Lada 2001; Bacmann et al. 2000; Ciolek \& Basu 2000; Zucconi, Walmsley, \& Galli 2001). 
It is also consistent with the observed near-virial-equilibrium of most
cores (Myers \& Goodman 1988).

We also note that the Bok globules, which are by definition
isolated sites of star formation, have shapes that are not significantly 
different from that of 
molecular cloud cores and condensations embedded within larger clouds.
This suggests that the
environment in which the cores and condensations are embedded plays a relatively
insignificant role in their dynamics, i.e., the external pressure from the
parental cloud does not seem to be important at this stage.

The smallest objects in our study, the condensations mapped in millimeter
and submillimeter continuum emission, may be the precursors to individual 
stars since
the mass spectrum appears to match the Initial Mass Function (IMF) compiled
by Salpeter (1955) over a certain mass range (Motte et al. 1998).
The estimated triaxial but near-oblate shape of these objects 
are an important link in
understanding the collapse process that leads to star formation.

For the larger molecular cloud scale, we have utilized an exhaustive
catalogue of the shapes of clouds in the outer Galaxy (Heyer et al. 2001).
Although our study of molecular cloud shapes is based on this single available
sample of projected axis ratios, and the result should be confirmed when
other shape data become available, the sheer size 
of this catalogue is a strong point. 
The histogram of observed axis ratios (Fig. \ref{heyertotalbest}) is very distinct 
from any of the other samples. It has a very sharp peak and
a severe lack of objects with $p\gtrsim 0.5$. While there may be some unknown
selection effect which biases against the observation of near-circular objects,
we note that the 
the orientations of the projected shapes in the plane of the sky
do appear to be truly random. 
Furthermore, we note that the earlier $^{13}$CO catalogue
of only 23 clouds in Ophiuchus by Nozawa et al. (1991) that was utilized by
Ryden (1996) has the same qualitative feature of a narrow peak near 
$p=0.3$ and a steep decline toward $p=1$.

Heyer et al. (2001) note that the vast majority of their clouds (all but the
largest clouds, which we loosely label GMC's) are not self-gravitating. 
They are either 
transient features or are held together by external pressure. If these
clouds are indeed brought together by large scale turbulence in the 
interstellar medium (or even confined for some time by an anisotropic 
ram pressure) we might expect that they have an elongated, filamentary 
shape (see e.g., Nagai, Inutsuka, \& Miyama 1998; Balsara, Ward-Thompson, 
\& Crutcher 2001; review by Shu et al. 1999). 
Since even the largest clouds seem to have these
shapes, we surmise that all clouds may be brought together by 
external forcing (due to shock waves or turbulent motions for example),
with only the largest clouds or densest regions within
smaller clouds able to become self-gravitating. 
This ties in with the
general picture of a rapid formation of molecular clouds due to 
external triggers (see e.g., Hartmann, Ballesteros-Paredes, \& Bergin 2001; 
Pringle, Allen, \& Lubow 2001).

\section{Summary}

Generally speaking, the observed decline in the observed probability
distribution function $\phi(p)$ towards $p=1$ favors triaxial rather
than axisymmetric intrinsic shapes. In addition, objects observed on
scales few $\times \, 0.1$ pc and smaller have a broad peak and a significant
number of objects observed with $p \approx 1$. This favors near-oblate
triaxial objects, as shown in detail in Paper I.
In contrast, molecular clouds observed on scales $\gtrsim 1$ pc have an
observed $\phi(p)$ with a much sharper peak and a precipitous decline
toward $p=1$. This favors near-prolate triaxial objects. Reviews of
the expected distributions $\phi(p)$ for various shapes can
be found in Binney \& Merrifield (1998) and Paper I.
A summary of our best fits to the various data sets is given in 
Table~\ref{summary}.

Our new results strengthen the finding that
one of the best fit intrinsic axis ratios is 
always quite a bit larger than the other axis ratio for cores, 
condensations, and Bok globules,
which means that these objects are preferentially flattened in one 
direction, and close to oblateness. They may then not be particularly
far removed from equilibrium or from oblate models often used to fit them.
On the other hand, the much larger, lower density clouds have best fit 
distributions such that many objects are 
close to prolateness, consistent with the formation of these objects due to
large scale external forcing. 

\begin{acknowledgements}

We thank John Dubinski for supplying the Monte Carlo program, as well as
the anonymous referee and scientific editor, Steven Shore whose
comments and careful reading helped to improve the paper.
This research was supported by a grant from NSERC, the Natural 
Sciences and Engineering Research Council of Canada. C. E. J. acknowledges
financial support from an NSERC postdoctoral fellowship.

\end{acknowledgements}

\end{document}